\documentclass[fleqn,10pt]{wlscirep}
\usepackage[utf8]{inputenc}
\usepackage[T1]{fontenc}
\usepackage{amsmath}
\usepackage{amssymb}
\usepackage{graphicx}
\usepackage{dcolumn}
\usepackage{bm}

\title{\textbf{Rotation related systematic effects in a cold atom interferometer onboard a Nadir pointing satellite}}

\author[1,*]{Quentin Beaufils}
\author[2]{Julien Lefebve}
\author[1]{Joel Gomes Baptista}
\author[1]{Raphaël Piccon}
\author[1]{Valentin Cambier}
\author[1]{Leonid A. Sidorenkov}
\author[2]{Christine Fallet}
\author[2]{Thomas L\'ev\`eque}
\author[1]{Franck Pereira Dos Santos}

\affil[1]{LNE--SYRTE, Observatoire de Paris, Universit{\'e} PSL, CNRS:UMR 8630, Sorbonne Universit{\'e}, 61 avenue de l'Observatoire, F--75014 Paris, France}
\affil[2]{Centre National d’Etudes Spatiales, 18 avenue Edouard Belin, 31400 Toulouse, France}

\affil[*]{quentin.beaufils@obspm.fr}

\begin{abstract}
We study the effects of rotations on a cold atom accelerometer onboard a Nadir pointing satellite. A simulation of the satellite attitude combined with a calculation of the phase of the cold atom interferometer allow us to evaluate the noise and bias induced by rotations. In particular, we evaluate the effects associated to the active compensation of the rotation due to Nadir pointing. This study was realized in the context of the preliminary study phase of the CARIOQA Quantum Pathfinder Mission.
\end{abstract}
\begin{document}

\flushbottom
\maketitle
%
%
\thispagestyle{empty}

\section{Introduction}

Quantum inertial sensors based on cold atom interferometry have reached a level of performances and maturity allowing for scientific and commercial use in various ground based applications \cite{Geiger2020}. Their sensitivity and accuracy are expected to increase dramatically in microgravity, where the interrogation time is no longer limited by the size of the instrument \cite{microgravity,Frye2021}. This triggered a recent interest in developing this technology for space \cite{Kaltenbaek2021,Alonso2022,Chiow2015} with potential applications in geodesy \cite{Carraz2014,leveque2021,Trimeche2019,Migliaccio2019,Zahzam2022}, fundamental physics \cite{Aguilera2014,Bassi2022}, navigation and gravitational wave observation \cite{Dimopoulos2008,Hogan2011}. Opening the way to those developments, the CARIOQA Pathfinder Mission \cite{CARIOQA} aims at realizing the first quantum accelerometer on a satellite. This space mission will consist in a single axis cold atom accelerometer designed to measure the non-gravitational acceleration along the velocity direction of a satellite on low Earth orbit. The mission is meant to realise key milestones for space atom interferometry and demonstrate unprecedented performances for quantum accelerometers.

Adapting cold atom accelerometers to low orbit environment requires to take into account residual fluctuating rotation of the satellite due to atmospheric and solar winds. Besides, various future missions (including CARIOQA) will require to operate in a Nadir pointing orbit, which implies a constant angular velocity in the mrad$/$s range. It has been identified \cite{Lan2012} that this rotation must be compensated to avoid a dramatic loss of contrast due to inhomogeneous Coriolis effect. In this article, we study the impact of rotation on the sensitivity and accuracy of the future CARIOQA cold atom accelerometer. In section II, we evaluate the phase of a rotating atom interferometer, with and without compensation, and identify a systematic effect related to rotation compensation. This systematic effect is then demonstrated experimentally. In section III, we realize a simulation of the attitude of a low orbit satellite to obtain angular velocity time series. In section IV, we use those time series to estimate the noise associated to residual rotation fluctuations and to the rotation compensation system.

\section{Evaluation of the phase of a rotating atom interferometer}

\subsection{General case}

\begin{figure}[h]
\centering
\includegraphics[width=0.6\linewidth]{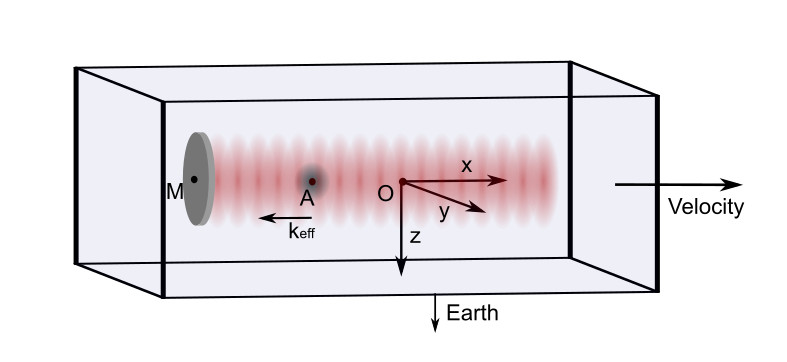}
\caption{\label{fig:satellite} Scheme of the interferometer geometry in the satellite. The reference frame used in this work has the satellite center of mass O as origin, the flight vector axis as x, the cross track axis, perpendicular to orbital plane, as y, and the Nadir axis as z. The atom cloud's mean position is A and the mirror's position is M.}
\end{figure}

The system considered is a single axis Mach Zehnder-like Chu Bordé cold atom interferometer \cite{Peters1999,mcguirk2002,louchet2011} embarked in a Nadir pointing satellite. The interferometer's measurement axis is aligned with the velocity vector axis (x axis). A mirror (M) used to retro-reflect the interferometer laser beams is rigidly fixed to the case of the satellite and constitutes the reference against which the acceleration of the free falling atom cloud (A) is measured. In this configuration, the instrument measures the non gravitational acceleration of the satellite along the x axis. All positions are defined relative to the center of mass of the satellite O which is taken as the origin of a reference frame aligned with the satellite main axes (see figure \ref{fig:satellite}).

We evaluate the output phase of the interferometer in the presence of an acceleration along x and a small angular velocity $\boldsymbol{\Omega}(t)=(\Omega_{x}(t),\Omega_{y}(t),\Omega_{z}(t))$ on O ($\Omega T \ll 1$ where T is the time between two interferometer's pulses, and $2T$ is the total duration of the interferometer). The configuration considered is a three pulses Raman interferometer with double diffraction \cite{leveque2009}. The total output phase difference can be expressed as a sum of the laser interaction phases during the interferometer : 

\begin{equation}\label{eq:Deltaphi1}
\Delta\phi = 2 \phi_{1} -2\phi_{A2}-2\phi_{B2}+\phi_{A3}+\phi_{B3},
\end{equation}
where the letters label the two paths of the interferometer and the numbers the three successive pulses. 

The effective laser wave vector $\boldsymbol{k}_{eff}$ is defined by the orientation of the retroreflection mirror M. When the mirror is fixed in the satellite frame, $\boldsymbol{k}_{eff}$ is constant and we can write :

\begin{equation}\label{eq:Deltaphi2}
\Delta\phi = 2 \textbf{k}_{eff} \cdot[\textbf{r}_{0} - 2( \textbf{r}_{A}(T)+\textbf{r}_{B}(T)) +( \textbf{r}_{A}(2T)+\textbf{r}_{B}(2T))],
\end{equation}

where $\textbf{r}_{A}(t)$ (resp. $\textbf{r}_{B}(t)$) is the classical position of the atomic wave packet on the path A (resp. B) of the interferometer at time t. 

We evaluate the classical trajectory of the atoms in the non Galilean satellite reference frame by solving the equations of motion using a polynomial ansatz with the Lagrangian : 

\begin{equation}
    L=\frac{1}{2} m(\dot{\boldsymbol{r}} + \boldsymbol {\Omega} \times \boldsymbol{r})^{2} + m \boldsymbol{a} \cdot \boldsymbol{r}
\end{equation}

where $\boldsymbol{a}$ is the linear acceleration of the atoms relative to the satellite case, which corresponds to the non gravitational acceleration of the satellite, and $m$ is the atomic mass. The terms due to gravity gradient are omitted as they are out of the scope of this article. 

Atomic wave packet positions are evaluated at each pulse and inserted in equation \ref{eq:Deltaphi1}. We obtain to second order in $\Omega T$:

\begin{align}
\Delta\phi = 2 k_{eff} T^{2} &[  a_{x}\\
& + 2 v_{z0}\Omega_{y} + 2 v_{y0}\Omega_{z}\label{coriolis1}\\
& + z_{0} \Omega_{x} \Omega_{z} - x_{0} (\Omega_{z}^{2}+\Omega_{y}^{2}) + y_{0} \Omega_{y} \Omega_{x}]\label{centrifugal1}\, ,
\end{align}

where $\textbf{r}_{0}=(x_{0},y_{0},z_{0})$ is the initial position of the atoms and $(v_{x0},v_{y0},v_{z0})$ their initial velocity in the satellite frame. Here $\boldsymbol{\Omega}=(\Omega_{x},\Omega_{y},\Omega_{z})$ is the temporal mean of the time dependant angular velocity during the interferometer. Besides acceleration along the x axis, this expression shows that the interferometer is sensitive to Coriolis (term \ref{coriolis1}) and centrifugal accelerations (term \ref{centrifugal1}). 

\subsection{Effect of a rotation compensation}\label{section2b}

In Nadir pointing navigation at the considered altitude, the satellite has a mean angular velocity of $\Omega_{N}\simeq 1$ mrad$/$s around the y axis. Given the initial velocity spread of the atoms, Coriolis acceleration is expected to induce a phase inhomogeneity across the cloud and limit the instrument \cite{Lan2012}. Even at a temperature as low as $T_{e}\simeq 100$ pK, the interferometer contrast would drop drastically for interrogation times $T$ larger than 1 s. 

The commonly accepted solution is to compensate the Nadir rotation by counter-rotating the retro-reflection mirror at the same rate during the interferometer \cite{Freier2016,Dickerson2013}. The effective wave vector $\boldsymbol{k}_{eff}$ has then a fixed direction in the inertial frame but varies in the satellite frame. In this case equation \ref{eq:Deltaphi2} is no longer valid.

To take into account the variation of $\boldsymbol{k}_{eff}$ in the satellite frame during a rotation compensated interferometer sequence, we define $\boldsymbol{\Omega}_{M}=(0,\Omega_{M},0)$ as the constant angular velocity of the mirror, and $\boldsymbol{r}_{M}=(x_{M},y_{M},z_{M})$ the position of its center of rotation. We also consider the possibility to rotate the incoming laser beam and define its angular velocity $\boldsymbol{\Omega}_{I}=(0,\Omega_{I},0)$. 

\begin{figure}[h]
\centering
\includegraphics[width=0.6\linewidth]{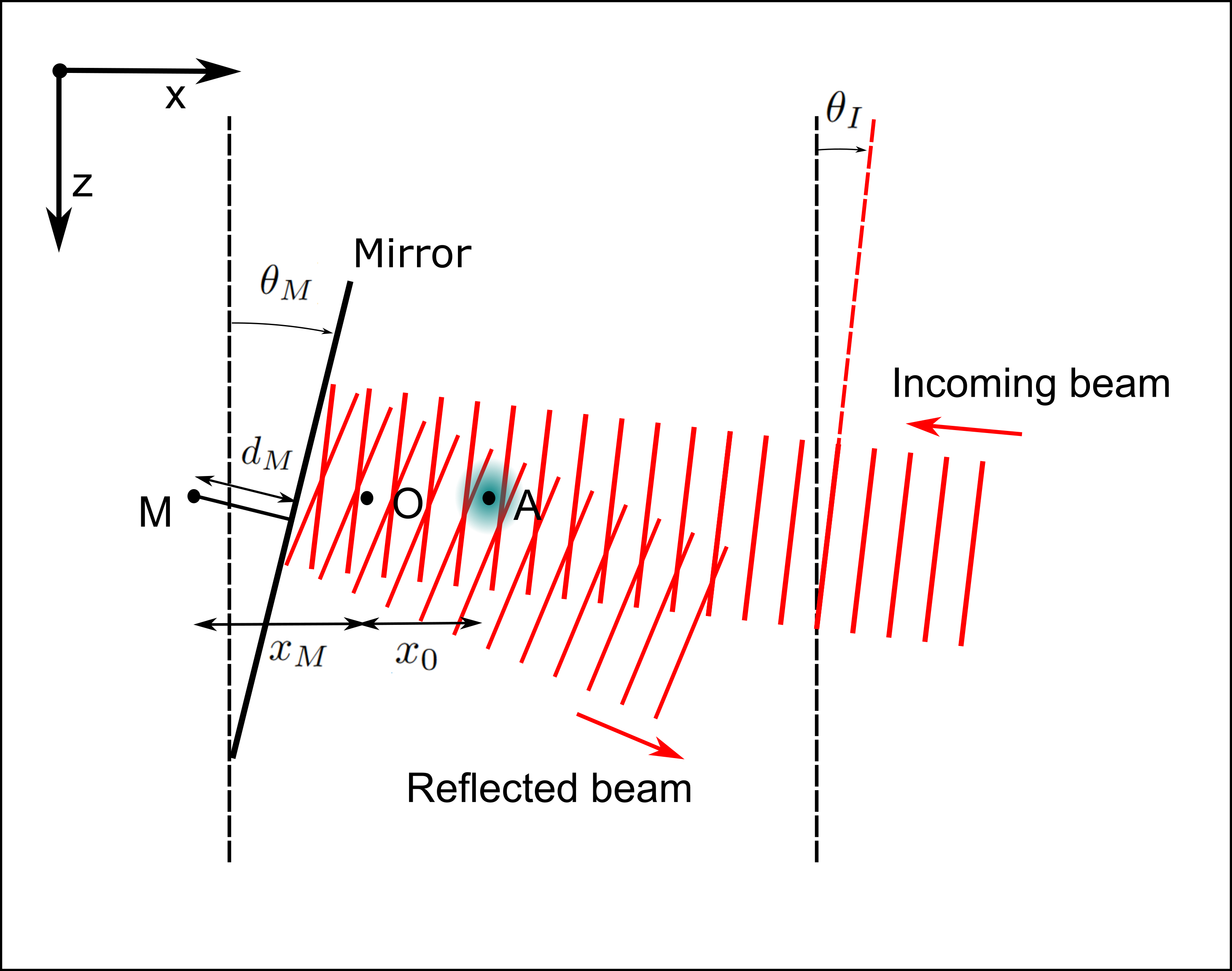}
\caption{Outline of  the geometrical configuration of the rotation compensated atom interferometer. The atomic cloud (A) is interrogated by a Raman beam with incoming angle $\theta_{I}$, reflected on a mirror rotating around the point M, and forming an angle $\theta_{M}$ with the z axis. \label{fig:angles}}
\end{figure}

A geometrical calculation of the laser phase at position $\textbf{r}=(x,y,z)$ as a function of the mirror and incoming beam orientation yields :

\begin{align}\label{eq:rotPhi}
    \phi(\boldsymbol{r},\boldsymbol{r}_{M},\theta_{I},\theta_{M})=&k_{eff} \cos(\theta_{I}-\theta_{M}) [(x_{M}-x)\cos(\theta_{M})
    -(z_{M}+z)\sin(\theta_{M})-d_{M}] ,
\end{align}
where we define $d_{M}$ as the distance between the mirror's reflection plane and its center of rotation $M$. $\theta_{M}$ (resp. $\theta_{I}$) is the angle between the mirror's reflection plane (resp. the incoming beam's equiphase plane) and the z axis (see figure \ref{fig:angles}). For the following, both angles are linked to angular velocities by the relations :

\begin{align}
&\theta_{i}(t=0)=-\Omega_{i} T\nonumber\\
&\theta_{i}(T)=0\nonumber\\
&\theta_{i}(2T)=\Omega_{i} T\, ,
\end{align}

where $i$ stands for $M$ and $I$. 

Using expression \ref{eq:rotPhi} in the semi-classical description detailed in section A, we obtain :

\begin{align}
\Delta\phi = 2 k_{eff} T^{2} &[  a_{x}\nonumber\\
& + 2 v_{z0}(\Omega_{y}+\Omega_{M}) + 2 v_{y0}\Omega_{z}\label{coriolis2}\\
& + z_{0} \Omega_{x} \Omega_{z} - x_{0} (\Omega_{z}^{2}+\Omega_{y}^{2}) + y_{0} \Omega_{y} \Omega_{x}\label{centrifugal2}\\
&+(x_{0}-x_{M})(\Omega_{M}^{2}+(\Omega_{M}-\Omega_{I})^{2})]\label{geometrical}\, .
\end{align}

As expected, the Coriolis acceleration due to $\Omega_{y}$ vanishes for $\Omega_{M}=-\Omega_{y}$. Besides, a term (\ref{geometrical}) corresponding to the motion of the mirror in the satellite appears in the phase expression. The centrifugal acceleration due to Nadir rotation around y axis is compensated only when this term is equal to $x_{0}\Omega_{y}^{2}$, which requires  $\Omega_{M}=\Omega_{I}=-\Omega_{y}$, but also $x_{M} = 0$. This corresponds to the situation when the mirror's rotation axis is aligned with the satellite's inertial center, which is the only situation when the mirror's reflection plane doesn't move in the inertial frame.

We observed experimentally this phase term (\ref{geometrical})  using a ground based cold atom gravity gradiometer. The experimental setup, described in \cite{Caldani2019}, consists in two cold atom clouds in free fall, vertically separated by a distance $d=1$ m, and interrogated by the same pair of counter-propagating Bragg laser beams. The common reference mirror is mounted on a tip-tilt platform and can be rotated with a controlled constant angular velocity $\Omega_{M}$ during the interferometer sequence. The incoming laser beam was fixed ($\Omega_{I}=0$), and for a differential measurement the rotation compensation phase can be expressed as : $$\Delta\phi_{RC}/(k_{eff} T^{2})= 2d\Omega_{M}^{2}.$$

\begin{figure}[h]
\centering
\includegraphics[width=0.9\linewidth]{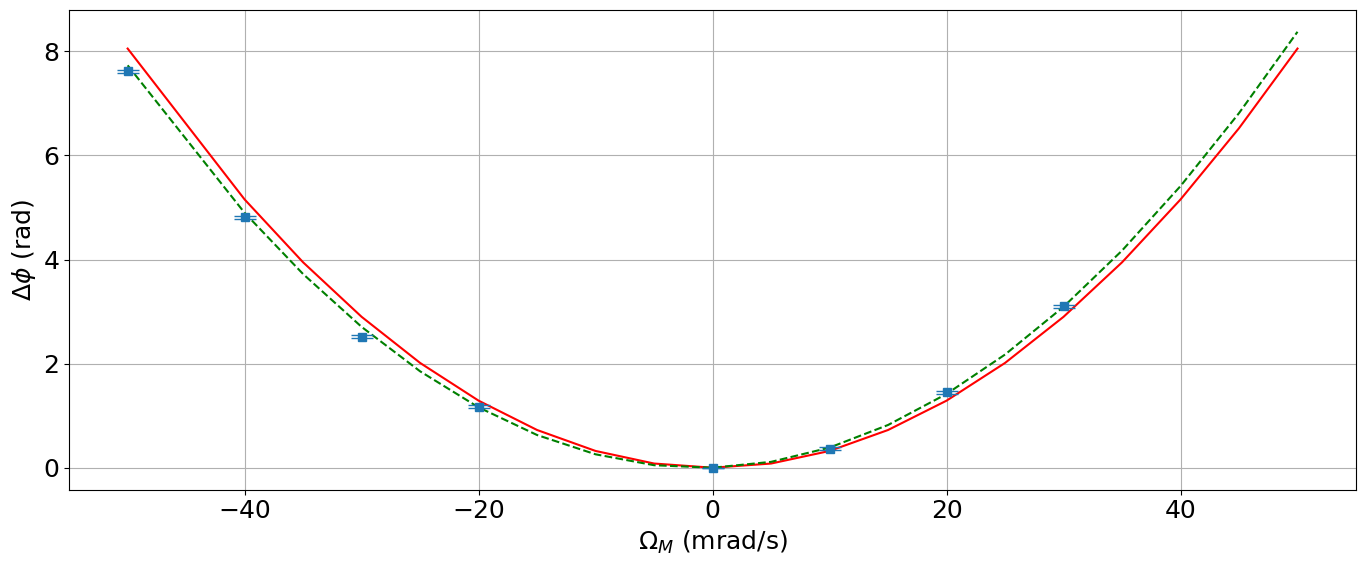}
\caption{ Differential phase of the vertical gravity gradiometer, as a function of the reference mirror's angular velocity during the measurement. Red solid line :  quadratic term $\Delta\phi_{RC}=4 k_{eff} T^{2}d\Omega_{M}^{2}$. Green dashed line : fit of the data with $f(\Omega_{M})=2 k_{eff} T^{2}(d\Omega_{M}^{2}+\delta v_{0}\Omega_{M})$. \label{fig:experiment}}
\end{figure}

Here we use single diffraction Bragg pulses, which explains a phase smaller by a factor of 2 compared to the calculation for double diffraction. In order to probe this result, we performed a differential phase measurement of the two interferometers as a function of the angular velocity. The differential phase at $\Omega_{M}=0$ was taken as a zero reference point in order to extract the angular velocity dependence only. The measurement is displayed in figure \ref{fig:experiment} and compared to theory without adjusted parameters (red solid line). We observed a small discrepancy, that can be explained by the presence of a residual differential transverse velocity $\delta v_{0}$ between the two clouds that causes a Coriolis effect $\Delta\phi_{coriolis}/(k_{eff} T^{2})= 2\delta v_{0}\Omega_{M}$. The green dashed line shows the result of a fit of the data with the sum of the quadratic and linear term, using $\delta v_{0}$ as the only adjustable parameter. We obtain the value $\delta v_{0}=1.3 \pm 0.11$ mm$/$s. Residual dispersion of the measured phase around the expected value larger than statistical uncertainty can be due to laser phase profile inhomogeneity.

\section{Simulation of the satellite attitude}

\begin{figure}
\centering
\includegraphics[width=1\linewidth]{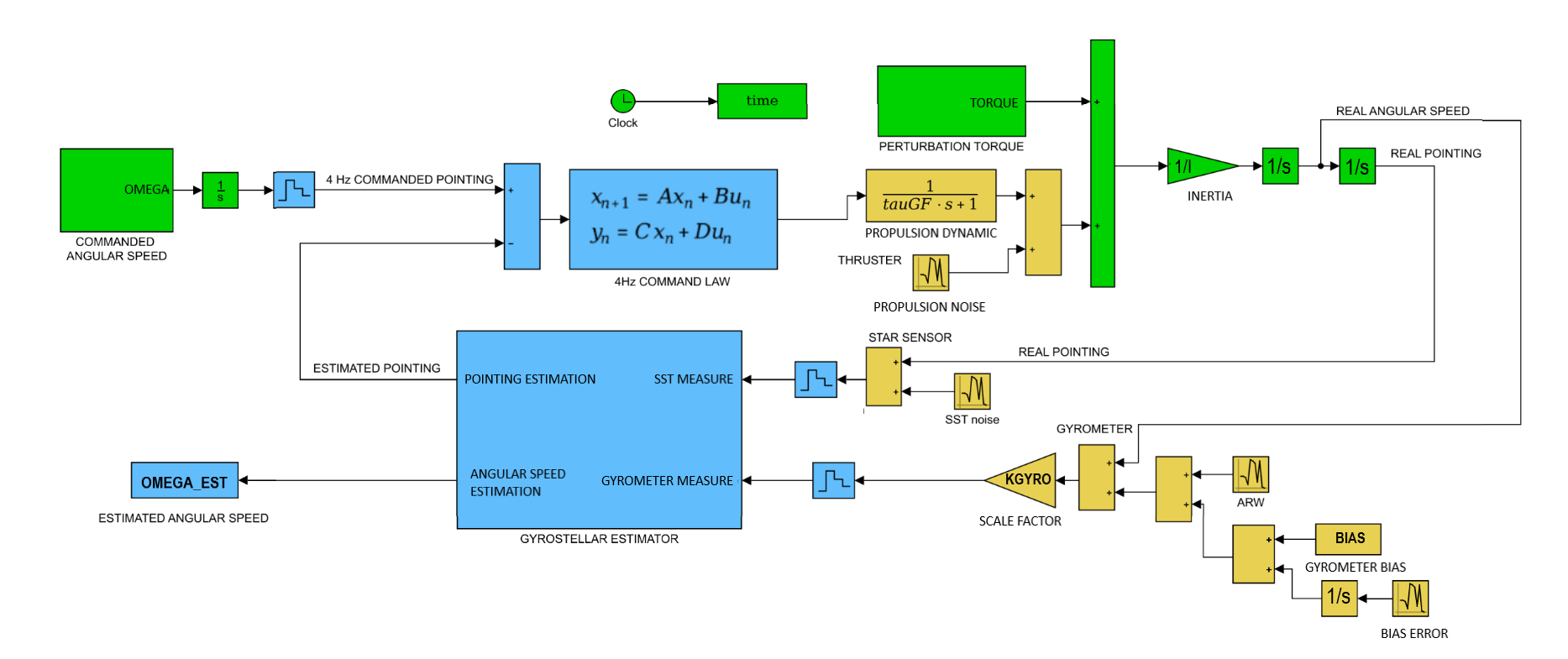}
\caption{Diagram of the AOCS simulator. The dynamics and perturbation torques are in green, the equipment models are in yellow and the software parts (control and
estimator) are in blue. \label{fig:AOCS}}
\end{figure}

We performed a Matlab/Simulink simulation of the future CARIOQA satellite attitude in order to obtain realistic angular velocity time series. Figure \ref{fig:AOCS} shows a diagram of the Attitude and Orbit Control System (AOCS) mono-axial simulator (each axis was simulated independently). The perturbation torque on the satellite was computed using a complex simulator able to evaluate precise atmospheric, solar pressure, magnetic and gravity gradient torques. The actuator considered was an electric propulsion system modelized as a first order transfer function with a white noise consistent with the propulsion system used on the MICROSCOPE project \cite{touboul_microscope_2001}. We used a 4 Hz gyrostellar attitude determination system \cite{ghezal_gyro_2006}, composed of a star tracker consistent with the HYDRA equipement caracteristics, and a gyrometer consistent with the ASTRIX-NS. It consists in a simplified version of a gyrostellar Kalman filter with a low-pass second order filter on the star tracker measurements and a high-pass second order filter on the gyrometer measurements with the same cut-off frequency (the hybridization frequency). An estimation of the gyrometer default (bias and scale factor) was also added to the estimator part to improve the angular velocity estimation. 

\begin{figure}[h]
\centering
\includegraphics[width=0.8\linewidth]{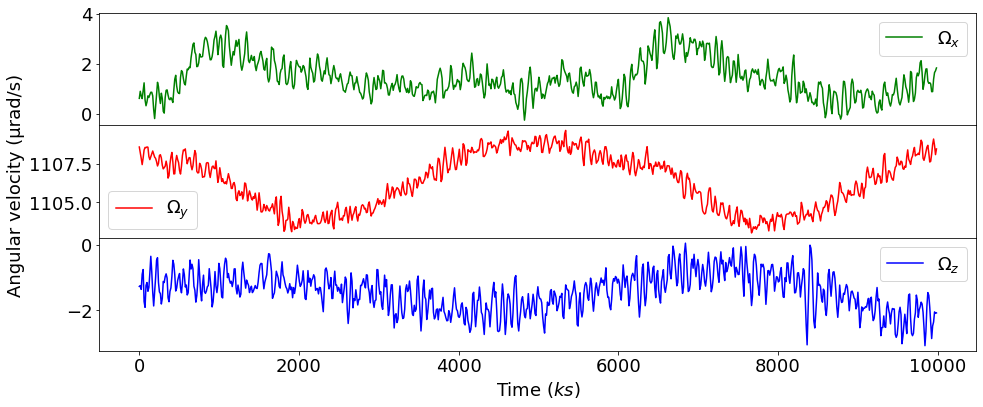}
\caption{Angular velocity time series of the satellite around the x axis (green), the y axis (red) and the z axis (blue).  \label{fig:rotations}}
\end{figure}
 
 The resulting angular velocity time series are presented in figure \ref{fig:rotations}. Residual angular velocity fluctuations are at the $\mu$rad$/$s level. A periodic variation of the commanded angular velocity around y axis with an amplitude of about $3$ $\mu$rad$/$s was added to take into account the orbit eccentricity. 

\section{Results and discussion}
\subsection{The inhomogeneous Coriolis effect}

\begin{figure}[h]
\centering
\includegraphics[width=0.5\linewidth]{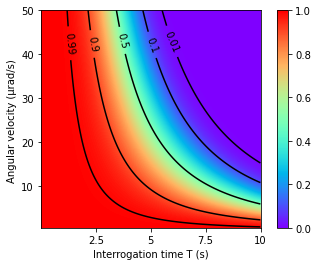}
\caption{Inhomogeneous Coriolis acceleration limited contrast of the interferometer as a function of the total angular velocity and interrogation time, for an atomic effective temperature of $\Theta = 4 \times 10^{-11} $ K. \label{fig:Contrast2D}}
\end{figure}

The velocity spread of a finite temperature atom cloud induces a phase inhomogeneity that alters the contrast of the interferometer \cite{Lan2012}. The CARIOQA interferometer will rely on a delta kick collimated atom source similar to the one described in \cite{deppner_collective-mode_2021}, which effective temperature can be lowered to $\Theta = 4 \times 10^{-11} $ K. Figure \ref{fig:Contrast2D} shows the contrast at this temperature, as a function of the interrogation time and the angular velocity's absolute value. We observe that the contrast vanishes rapidly when $T$ increases, unless the residual angular velocity remains close to $1$ $\mu$rad$/$s. This can be achieved by counter rotating the reference mirror, as shown in section \ref{section2b}. This implies a real time knowledge of the satellite's angular velocity during the interferometer sequence to that level of stability, but also a control over the angle of the rotating mirror and incoming beam down to about a few microrad. 

For the following, we consider a rotation compensation with a shot to shot standard deviation of the angular velocity of $1$ $\mu$rad$/$s, with an angular velocity compensation perfectly aligned around the y axis, that follows the driven orbital angular velocity from the AOCS.

\subsection{Phase noise associated to residual rotation }

Using the phase expression found in section II.B and the attitude simulation described in section III, we estimate the phase noise associated with residual (uncompensated) rotation noise of the satellite and compare it to the expected sensitivity of a quantum projection noise (QPN) limited atom interferometer with $N=10^{5}$ atom, as expected for the CARIOQA instrument. The parameters used for this comparison are given in table \ref{tab:positionspec}. We use state of the art values for the interferometer, except for the interrogation time $T$ which is the main variable parameter for the future mission. As the future performances depend strongly on $T$, we consider different values between $1$ s and $5$ s.

\begin{table}[h]
\centering

\begin{tabular}{|c|c|c|c|}
\hline
\textrm{Parameter}&
\textrm{Value}&
\textrm{Unit}&
\textrm{Comment}\\
\hline

$T$ & $1$ to $5$ & s & Interferometer interrogation time\\
\hline
$N$ & $10^{5}$ & & Atom number \\
\hline
$C$ & $0.8$ & & Contrast \\
\hline
$r_{0}$ & $1$  & mm & Initial mean position of the cloud\\
\hline
$v_{0}$ & $100$  & $\mu$m/s & Initial mean velocity of the cloud\\
\hline
$\Theta$ & $40$  & pK & Effective temperature of the cloud\\
\hline

\end{tabular}
\caption{\label{tab:positionspec}Simulation parameters}
\end{table}

The Centrifugal acceleration noise depends on the initial distance between the atom cloud and the center of mass of the satellite. The control over this value is expected to be limited by the knowledge of the center of mass's position in the satellite frame, to about $1$ mm. The Coriolis acceleration depends on the initial mean velocity of the atom cloud. From \cite{deppner_collective-mode_2021}, we anticipate a delta kick collimated atom source with a residual initial mean velocity $v_{0}$ of about $100$ $\mu$m$/$s.

The effect of time-dependent perturbations on the atom interferometer is modelled using the formalism of the sensitivity function \cite{Cheinet2008}. If we describe the rotation noise source by a power spectral density (PSD) $S(f)$, the resulting RMS phase noise of the atom interferometer is calculated as : $$\sigma^{2}_{\phi}=\int_{0}^{\infty} S(f)| H(f) |^{2} \,df $$ where $H(f)$ is the atom interferometer transfer function. For this simulation, we also took into account a potential loss of contrast due to the inhomogeneous Coriolis acceleration.

\begin{figure}[h]
\centering
\includegraphics[width=0.8\linewidth]{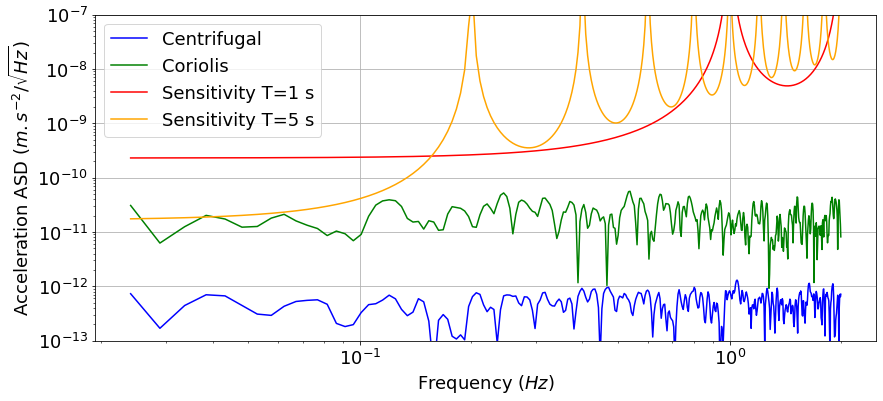}
\caption{\label{fig:coriolis}Amplitude spectral density of acceleration noise due to residual rotations of the satellite compared to expected QPN limited sensitivities calculated for $T=1$ s and $T=5$ s.}
\end{figure}

Figure \ref{fig:coriolis} shows the amplitude spectral density (ASD) of the Coriolis and centrifugal acceleration noise, as well as the expected QPN limited sensitivity of the instrument. We observe that maintaining a rotation induced phase noise level below QPN is thus possible using a state of the art atom interferometer and a commercial AOCS system.



\subsection{Phase noise associated to the rotation compensation}

\begin{figure}[h]
\centering
\includegraphics[width=0.8\linewidth]{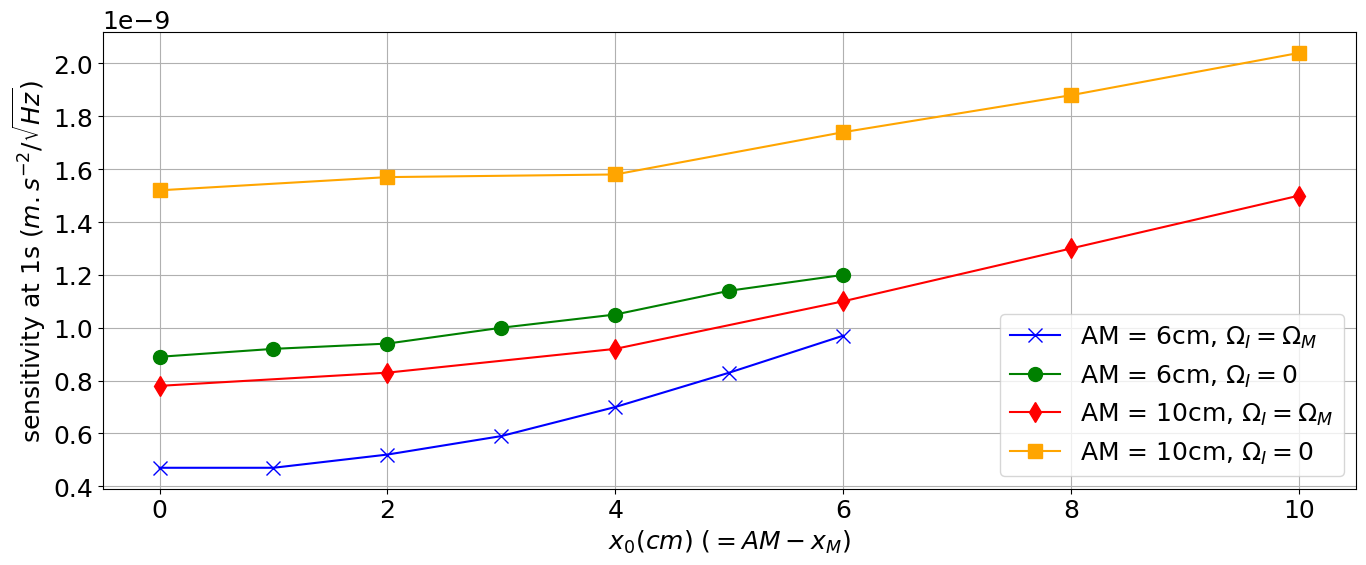}
\caption{Sensitivity limit associated to rotation noise, as a function of the position of the instrument (atoms and mirror) in the satellite, for two different distances between the atoms and the mirror's rotation center, and with or without incoming beam rotation. $x_{0}$ is the distance atoms - center of mass of the satellite.  \label{fig:SensitivityVScom}}
\end{figure}

As shown in section I, the nadir rotation compensation induces a phase shift to the interferometer if the center of rotation of the mirror ($M$) is not perfectly superimposed to the center of mass ($O$) of the satellite. A noise associated to this phase can appear because of an imperfect control of the rotating mirror's angle and limit the sensitivity of the instrument. The usually considered technical solution for rotating the mirror is to use a piezoelectric tip-tilt platform. With this technique $M$ is usually close to the mirror, which implies a distance of a few tens of centimeters between $M$ and the initial position of atoms $A$. Based on commercially available systems, we anticipate a shot to shot jitter of the mean rotation rate $\Omega_{M}$ of $\sigma_{\Omega_{M}}=1$ $\mu$rad$/$s. Under those circumstances a trade off appears between minimizing this noise and the centrifugal acceleration noise mentioned in section IV B, by respectively placing the center of mass of the satellite close to $M$ or close to $A$. This is illustrated by figure \ref{fig:SensitivityVScom} where we evaluated the rotation noise limited sensitivity, defined as the total phase noise of the instrument for an equivalent $1$ s measurement, as a function of the position of the atoms-mirror ensemble relatively to the center of mass ($x_{0}=OA$), for two different distances between the atoms and the mirror ($AM$). The smallest possible distance for a $t=5$ s interrogation time is $AM=6$ cm. We also compare between the situation when only the reference mirror is rotated ($\Omega_{I}=0$, $\Omega_{M}=- \overline{\Omega}_{y}$), and both the mirror and the incoming beam are rotated ($\Omega_{I}=\Omega_{M}=- \overline{\Omega}_{y}$). The parameters given in table \ref{tab:positionspec} were used for this evaluation. The sensitivity is limited by centrifugal acceleration for large $x_{0}$, and by rotation compensation noise for small $x_{0}$ (large $x_{M}$).

As expected, the shorter distance $AM$ is more favorable. Rotating the incoming beam also allows to improve the sensitivity by a factor two (see equation \ref{geometrical}, section \ref{section2b}). In the best case, we observe that the sensitivity of the instrument is limited to approximately $5 \times 10^{-10}$ m.s$^{-2}.$Hz$^{-1/2}$, which is above QPN for an interrogation time of $5$ s. A way to improve this limitation could be to use a specifically designed rotation compensation system with a custom position of the rotation center far from the mirror. This way, the rotation center of the mirror and the mean atomic position could both be superimposed to the center of mass of the satellite.

 Besides the limitation on the sensitivity of the instrument, its accuracy and long term stability can be also limited by the rotation compensation system. The phase term associated to the rotation of the mirror and of the incoming beam can be as large as a few $10^{-7}$ m.s$^{-2}$ if $x_{M}$ is a few centimeters. Satellite navigation relies on rocket fuel propulsion that may cause a long term drift of the position of the center of mass. Under the conditions considered, the measurement bias associated to mirror rotation is on the order of $10^{-7}$ m.s$^{-2}$ with a variation of $10^{-9}$ m.s$^{-2}$ per millimeter drift of the position of the center of mass. A precise modelling of the center of mass variations may be necessary to evaluate this systematic effect. Alternatively, a calibration of the bias term could be performed by varying the satellite's total angular velocity.

\section{Conclusion}

In this article we evaluated the effect of rotations on the performances of a cold atom accelerometer in a low orbit Nadir pointing satellite for the future CARIOQA mission. A simulation of the satellite's attitude along with a calculation of the interferometer's phase allowed to specify some instrumental parameters such as the initial velocity of the atomic cloud, its positioning in the frame of the satellite and its temperature. We showed that it is possible to maintain the rotation related phase noise limit on the sensitivity of the instrument down to a few $10^{-10}$ m.s$^{-2}$.Hz$^{-1/2}$ in low orbit with a relatively standard attitude control system and state of the art ultra-cold atom technology. This was illustrated by a specific set of parameters, however future design studies may aim at better performances by either improving the AOCS or the cold atom system.

When the Nadir rotation is compensated, we identified a phase term due to an imperfect alignment between the rotation axis of the mirror and the center of mass of the satellite. This term can limit the sensitivity of the instrument, as well as its accuracy. The rotation compensation system has to be designed in accordance in the frame of the CARIOQA Quantum Pathfinder Mission, as well as in other future space atom interferometry missions.

\section*{Methods}

 The datasets used and/or analysed during the current study available from the corresponding author on reasonable request.

\bibliography{biblio}

\section*{Acknowledgements}

The authors thank Peter Wolf and Robin Corgier for stimulating discussions. 

\section*{Author contributions statement}

Q.B. and F.P. did the calculations,  J.G.B., V.C. and L.A.S. conducted the experiment, J.L., C.F. and T.L. conducted the attitude control simulation and Q.B. analysed the results.  All authors reviewed the manuscript.

\end{document}